\documentclass[preprint,showpacs,preprintnumbers,amsmath,amssymb,superscriptaddress, titlepage]{revtex4-2}
\usepackage{graphicx}
\usepackage{dcolumn}
\usepackage{bm}
\usepackage{hyperref}

\begin{document}

\title{Prospects of searching for unstable nucleus states in relativistic nuclear fragmentation}

\author{D.A. Aretemenkov}
\affiliation {Joint Institute for Nuclear Research, Dubna, Russia}

\author{V. Bradnova}
\affiliation {Joint Institute for Nuclear Research, Dubna, Russia}

\author{O.N Kashanskaya}
\affiliation {Gomel State University, Gomel, Republic of Belarus}

\author{N.V. Kondratieva}
\affiliation {Joint Institute for Nuclear Research, Dubna, Russia}

\author{N.K. Kornegrutsa}
\affiliation {Joint Institute for Nuclear Research, Dubna, Russia}

\author{E. Mitsova}
\affiliation {Joint Institute for Nuclear Research, Dubna, Russia}
\affiliation {Institute for Nuclear Research and Nuclear Energy, Sofia, Bulgaria}

\author{N.G. Peresadko}
\affiliation {Physical Institute of the Russian Academy of Sciences, Moscow, Russia}
 
\author{V.V. Rusakova}
\affiliation {Joint Institute for Nuclear Research, Dubna, Russia}

\author{R. Stanoeva}
\affiliation {Southwestern University, Blagoevgrad, Bulgaria}
\affiliation {Institute for Nuclear Research and Nuclear Energy, Sofia, Bulgaria}

\author{A.A. Zaitsev}\email{zaicev@lhe.jinr.ru}
\affiliation {Joint Institute for Nuclear Research, Dubna, Russia}
\affiliation {Physical Institute of the Russian Academy of Sciences, Moscow, Russia}

\author{P.I. Zarubin}
\affiliation {Joint Institute for Nuclear Research, Dubna, Russia}
\affiliation {Physical Institute of the Russian Academy of Sciences, Moscow, Russia}

\author{I.G. Zarubina}
\affiliation {Joint Institute for Nuclear Research, Dubna, Russia}

\begin{abstract}

The article is dedicated to the experimental study in the relativistic approach to the problems of nuclear cluster physics for the prospects of the \href{http://becquerel.jinr.ru/}{BECQUEREL} experiment. The nuclear emulsion method applied in this experiment makes it possible to study thoroughly the relativistic final states in the fragmentation of nuclei. The focus of the presented research is the dynamics of emergence of the $^{8}$Be nucleus and the Hoyle state, as well as the search for the 4$\alpha$-particle condensate decaying via the above nuclear states. In this context, the analysis of exposure to $^{84}$Kr nuclei at 950 MeV/nucleon is shown. As a continuation of the study of light nuclei, we have demonstrated the search for the isobar-analogue state of the $ ^{13} $N nucleus in the fragmentation of $ ^{14} $N nuclei at 2 GeV/nucleon.

\end{abstract}


\maketitle

\section*{Introduction}
\noindent The presence of quartets of spin-paired protons and neutrons in the structure of light nuclei manifests itself in the intense formation of $\alpha$-particles in a wide variety of nuclear reactions and decays \cite{Ajzenberg}. The study of ensembles consisting of several $\alpha$-particles makes it possible to reveal the role of the unstable $^{8}$Be and $^{9}$B nuclei and search for their analogs, starting from the Hoyle 3$\alpha$-state (HS). The $\alpha$-clustering is most pronounced in the $^{8}$Be nucleus. The decay energy $^{8}$Be $\to$ 2$\alpha$ is only 91.8 keV. Its width, 5.57 $\pm$ 0.25 eV, corresponds to the lifetime that is 8–9 orders of magnitude larger than the reaction time scale. $^{8}$Be is an essential decay product of $^{9}$B and HS. The ground state of the $^{9}$B nucleus is higher than the $^{8}$Be$p$ threshold by 185.1 keV, and its width, which is 0.54 $\pm$ 0.21 keV, also points to it as a long-lived state.

The HS state is the second excitation of the $^{12}$C nucleus (review \cite{Freer}) at 378 keV above the 3$\alpha$ threshold. The isolation of HS at the beginning of the $^{12}$C excitation spectrum and the width $\Gamma$(HS) = 9.3 $\pm$ 0.9 eV make it a 3$\alpha$ analogue of $^{8}$Be. Synthesis of $^{12}$C in the medium of red giants is possible through the merger 3$\alpha$ $\to$ $\alpha$$^{8}$Be $\to$ $^{12}$C(0$^{+}_{2}$) $\to$ $^{12}$C (+2$\gamma$ or $e^+e^-$ with a probability of about 10$^{-4}$). Further synthesis of $\alpha$$^{12}$C $\to$ $^{16}$O$\gamma$ through the $^{16}$O level of suitable energy is parity forbidden, which determines the relative abundance of $^{12}$C and $^{16}$O, and, in fact, the survival of $^{12}$C under the astrophysical conditions of helium burning. However, the synthesis of $^{16}$O is probable in the sequence $^{12}$C$^{12}$C $\to$ $^{12}$C$^{12}$C(0$^+_2$) $\to$ $^{16}$O$^{8}$Be \cite{Freer}.

Determining the key role of $^{8}$Be and HS in nuclear astrophysics, these facts assume an opportunity of the emergence of their heavier counterparts. Not limited to the role of excitation of the $^{12}$C nucleus, HS can also manifest itself in reactions with other nuclei, which combines it, like $^{8}$Be and $^{9}$B, with other fragments. The exotically large sizes of these three objects (for example, in \cite{Tohsaki}), predicted theoretically, are critical to understand the mechanism of their generation and fragmentation in general. Possessing a nuclear-molecular structure, they can serve as the founders of their own branches of excitations and states with a more complex composition.

The increased interest in the unstable states of $\alpha$-particles is motivated by the concept of $\alpha$-particle Bose-Einstein condensate ($\alpha$BEC), put forward in the early 2000s. in analogy with quantum gases of atomic physics (review \cite{Tohsaki}). Excitations of $n\alpha$-fold nuclei immediately above the binding thresholds of $\alpha$-particles can serve as manifestation of $\alpha$BEC. Coexisting with fermionic excitations, they are considered on the basis of the mean field of the bosonic type, formed by the gas of almost ideal bosons in the $S$-state at the average density four times lower than usual. $^{8}$Be and HS are described as 2- and 3$\alpha$BEC states, and their decays can serve as signatures for more complex $n\alpha$BEC decays. The existence of heavier analogues of HS may enrich nucleosynthesis scenarios on the way to heavy nuclei. Experimental approaches to search for $\alpha$BEC in the fragmentation of light nuclei have been proposed, among which the one is presented here (review \cite{Oertzen}. In focus is the 0$^+_6$ state of the $^{16}$O nucleus at 15.1 MeV (or 660 keV above the 4$\alpha$ threshold), considered as the 4$\alpha$ analog of HS with decay into $\alpha$HS or 2$^{8}$Be. The consideration of $n\alpha$BEC as weakly bound unstable states indicates new opportunities for their search with increasing energy and mass numbers of generating nuclei. It is very valuable to demonstrate the universality of $n\alpha$BEC candidates on the basis of relativistic invariance.

During fragmentation of relativistic nuclei, ensembles of He and H nuclei are generated in the extremely narrow cone. There are no thresholds to detect them, and energy losses are minimal. Because of the extremely low energy, the $^{8}$Be, $^{9}$B, and HS decays should manifest themselves as pairs and triplets of relativistic fragments of He and H with the smallest angles of expansion. According to the widths, the decays of $^{8}$Be, $^{9}$B and HS occur at ranges from several thousand ($^{8}$Be and HS) to several tens ($^{9}$B) of atomic sizes and should be identified by the minimum invariant mass. The answer to these challenges is provided by the nuclear emulsion (NE) method, whose application has been continued in the BECQUEREL experiment. In NE layers longitudinally exposed to relativistic nuclei, tracks of fragments are completely observed, and their directions are measured with the best resolution. Determination of the invariant masses of ensembles of relativistic fragments of He and H in the approximation of the conservation of the velocity of the parent nucleus makes it possible to project their angular correlations onto the relative energy scale, starting from the $^{8}$Be decay. The opportunities and status of these studies are presented in review publications \cite{Adachi,Zarubin,Artemenkov}. Among the achievements is identification of $^{8}$Be, $^{9}$B, as well as of the Hoyle state in the fragmentation of light nuclei, including the radioactive ones \cite{Zarubin}.

The BECQUEREL experiment was proposed to extend this approach to search for $\alpha$BEC states in fragmentation events of medium and heavy nuclei. Recently, a rapid increase in the contribution of $^{8}$Be, $^{9}$B, and HS with increasing in the number of accompanying $\alpha$-particles has been discovered. The explanation for this effect may lie in the picture of the combination of the formed $\alpha$-particles with increase in their density in the phase space \cite{Artemenkov}. The scenario suggests that $\alpha$BEC arises not as a result of suitable excitation of the parent nucleus, but as a result of forming $\alpha$BEC-type states by means of successive pickup of accompanying $\alpha$-particles. Then $\alpha$BEC can be regarded as a short-lived state of nuclear matter having extremely low density and temperature, not associated with the parent nucleus excitation. The selection of events with high multiplicity of $\alpha$-particles can be used as an amplifying factor in the statistics of events in $\alpha$BEC. Thus, the search for $\alpha$BEC based on the invariant mass of ensembles of relativistic $\alpha$-particles with extremely close 4-momenta is the nearest future for the BECQUEREL experiment, discussed below.

At the same time, the study of the unstable state formation by light nuclei will continue to search for isobar-analogue states (IAS) using the NE method in the relativistic approach. Responding due to mass effects of much higher energy and, at the same time, very small widths. In this aspect, the available training of nuclear power engineering with relativistic nuclei $^{14}$N, $^{22}$Ne, $^{24}$Mg, and $^{28}$Si is worth being analyzed. Currently, this search is underway for IAS $^{13}$N(15.065) in the $^{14}$N $\to$ 3$\alpha p$ fragmentation, IAS $^{8}$Be(16.6 + 16.9) in $^{9}$Be $\to$ 2$\alpha$, and IAS $^{9}$B(14.7) in $^{10}$C $\to$ 2$\alpha$2$p$. Its preliminary results are given below.

\section*{Opportunities of the NE method}
\noindent 
The opportunities of the NE method, which remains unique for the above measurements are to be reminded. The irradiated stacks are assembled from NE layers up to 10$\times$20 cm$^2$ in size, 200 $\mu$m on a glass substrate and 550 $\mu$m - without it. If the beam is directed parallel to the plane of the layers, then the tracks of all relativistic fragments remain long enough in one layer for 3D reconstruction of the angles. The substrate provides the track ``rigidity'', and its absence allows one to trace them to neighboring layers. The factors to obtain significant statistics are the thickness and the total solid angle of detection. The NE media contain close concentrations of Ag and Br and CNO atoms, as well as the threefold higher number of H. Scanning along the tracks of the nuclei under study on microscopes with 20$\times$ objectives provides detecting about a thousand of interactions without sampling or tens of the peripheral ones. The statistics of several hundred peripheral interactions with certain configurations of relativistic fragments is achievable with transverse scanning.

The tracks of relativistic He and H fragments visually identified by their charges are concentrated in a cone up to $sin\theta_{fr}$ = $p_{fr}$/$P_0$, where $p_{fr}$ = 0.2 GeV/$c$ is the quantity characterizing the Fermi momentum of nucleons in the projectile nucleus, and $P_0$ is its momentum per nucleon. Due to the grain size of about 0.5 $\mu$m, the angular resolution on the 1 mm base is not worse than 10$^{-3}$ rad. The transverse momentum of a $P_T$ fragment with mass number $A_{fr}$ is defined as $P_T$ $\approx$ $A_{fr}P_0sin\theta$ in the $P_0$ conservation approximation. The assignment of the mass numbers of H and He is possible from measurements of the mean angles of multiple scattering. The use of this time-consuming method is justified in special cases for a limited number of tracks. In the case of dissociation of stable nuclei, it is sufficient to assume that He corresponds to $^{4}$He and H - $^{1}$H. This simplification is especially true for the case of extremely narrow $^{8}$Be and $^{9}$B decays \cite{Zarubin}.

In the fragmentation of NE nuclei, the tracks of $b$-particles ($\alpha$-particles and protons with energies below 26 MeV), $g$-particles (protons with energies above 26 MeV), and also $s$-particles (produced mesons) can be observed. The approximate conservation of the charge of the beam nuclei in the event with a small number of slow fragments is used as a selection criterion for interactions of the peripheral type, which make up several percent of their total number of stars. The most peripheral interactions, called coherent dissociation or ``white'' stars, are not accompanied either by fragmentation of the target nuclei or meson production. On the site \href{http://becquerel.jinr.ru/}{BECQUEREL} photos and videos of characteristic interactions are available.

The invariant mass of any ensemble of relativistic fragments is defined as the sum of all products of 4-momenta $P_{i,k}$ of the fragments $M^{*2}$ = $\Sigma$($P_i\cdot P_k$). The subtraction of the mass of the initial nucleus or the sum of the masses of the fragments $Q$ = $M^*$ - $M$ is a matter of convenience of representation. The components $P_{i,k}$ are determined in the fragment conservation approximation $P_0$. The reconstruction from the invariant mass of the decays of the relativistic unstable nuclei $^{8}$Be and $^{9}$B, mastered in the BECQUEREL experiment, has confirmed the validity of this approximation \cite{Zarubin}.

The most accurate angle measurements has been obtained by using the coordinate method on KSM-1 microscopes (Carl Zeiss, Jena) with 60$\times$ objectives in immersion oil. The measurements are carried out in the Cartesian coordinate system. The NE layer is rotated so that the direction of the analyzed primary trace coincides with the OX axis of the microscope stage with a deviation of up to 0.1–0.2 $\mu$m per 1 mm of the track length. Then the OX axis of the system coincides with the direction of the primary projection onto the layer plane, and the OY axis on it is perpendicular to the primary track. The OZ axis is perpendicular to the plane of the layer. For OX and OY, the measurements are carried out by micro screws of the horizontal movement, and for OZ - by a depth-of-field micro screw. Coordinates are measured on the primary and secondary tracks from 1 to 4 mm long with a step of 100 $\mu$m. Planar and depth angles are calculated from their linear approximation.

\section*{Study status}
\noindent 
Basically, the search for $\alpha$BEC is carried out on the basis of compact spectrometers which provide a significant coverage of the solid angle in beams of light nuclei at several tens of MeV per nucleon \cite{Borderie,Barbui,Charity,Bishop,Smith,Manna,Adachi}. Placed in vacuum volumes near ultrathin targets the silicon detectors of the best energy resolution have been used in the experiments. Identification of the unstable nuclei and states is carried out by energy and angular correlations in the registered ensembles of $\alpha$-particles.

The experiment with full detection of $\alpha$-particle projectile fragments in the reaction $^{40}$Ca(25 MeV/nucleon) + $^{12}$C indicated an increasing contribution of $^{8}$Be up to the multiplicity of $\alpha$-particles equal to 6 \cite{Borderie}. This fact contradicts the model predicting its decrease (Table 2 in \cite{Borderie}). The search was made for the decays of the $^{16}$O(0$^+_6$, 15.1 MeV) $^{20}$Ne(12 MeV/nucleon) + $^{4}$He \cite{Barbui} and $^{16}$O(160, 280, 400 MeV) + $^{12}$C states \cite{Bishop}. Recently, the data on $^{16}$O(45 MeV) + $^{12}$C $\to$ 4$\alpha$ in full kinematics \cite{Manna} have been analyzed for all possible configurations. The excitation function was reconstructed 
directly from 4$\alpha$, as well as for special decay channels such as $^{12}$C(0$^+_2$)$\alpha$, $^{12}$C(3$^-_1$)$\alpha$, and 2$^{8}$Be. However, the search for 
the 15.1 MeV state has remained unsuccessful in all cases \cite{Smith}. Coincidence measurements were made for $\alpha$-particles (386 MeV) scattered at 0$\deg$ in the $^{20}$Ne($\alpha$,$\alpha '$)5$\alpha$ reaction \cite{Adachi}. It is stated that the newly observed states at 23.6, 21.8 and 21.2 MeV in $^{20}$Ne are strongly associated with the 4$\alpha$BEC $^{16}$O candidate being themselves the $\alpha$BEC candidates.

Although the status of $\alpha$BEC observations remains uncertain \cite{Bishop}, in all cases HS is produced not only while $^{12}$C fragmentation. This fact indicates independence on the parent nucleus of HS, as well as $^{8}$Be. The similar versatility should be shown by $\alpha$BEC candidates. In general, it seems that experiments to search for 4$\alpha$BEC states have reached a practical limit in terms of statistics. The orientation toward peripheral collisions of heavier nuclei with higher energy is required. To combine the data obtained in the widest possible energy range and, on this basis, to confirm the versatility of $\alpha$BEC, it is necessary to represent the unstable states in a relativistic invariant form. 

\begin{figure}[]
	\centerline{\includegraphics*[width=1.0\linewidth]{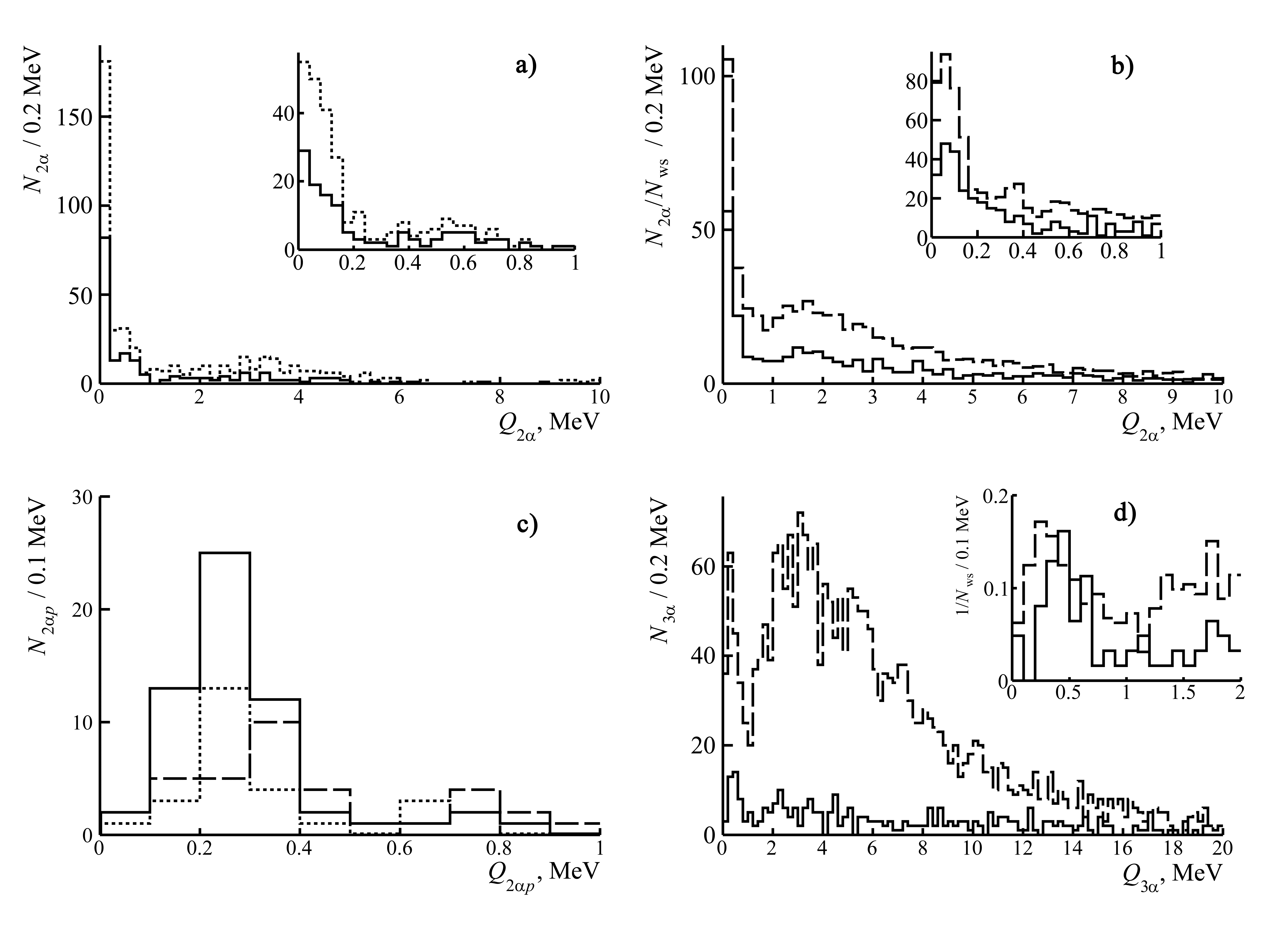}}
	\caption{Invariant mass distributions \cite{Artemenkov}: a) $Q_{2\alpha}$ in $^{9}$Be (1.2 GeV/nucleon) $\to$ 2$\alpha$ (dashed, solid – ``white'' stars; b) $Q_{2\alpha}$ in $^{12}$C(3.65 GeV/nucleon) $\to$ 3$\alpha$ (solid line) and $^{16}$O(3.65 GeV/nucleon) $\to$ 4$\alpha$ (dotted); c) $Q_{2\alpha p}$ ($<$ 1 MeV) into $^{10}$C(1.2 GeV/nucleon) $\to$ 2$\alpha$2$p$ (solid) and $^{11}$C(1.2 GeV/nucleon) $\to$ 2$\alpha$2$p$ (dots) and $^{10}$B(1 GeV/nucleon) → 2$\alpha p$ (dotted line); $Q_{3\alpha}$ in $^{12}$C(3.65 GeV/nucleon) $\to$ 3$\alpha$ (solid) and $^{16}$O(3.65 GeV/nucleon) $\to$ 4$\alpha$ (dashed).}
\end{figure}

Electronic experiments in beams of relativistic nuclei have not overcome the difficulties caused by the ionization quadratic dependence on charges, the extremely low divergence, and the coincidence of the magnetic rigidity of relativistic fragments and beam nuclei. The only practical alternative is to purposefully apply the technically simple and inexpensive NE method. It gives flexibility and uniformity at the search stage, and theoretically - transparency of interpretation. In the 70s, exposures of NE stacks to light nuclei of several GeV per nucleon started at the JINR Synchrophasotron and Bevalac LBL, and in the 90s - to medium and heavy at AGS (BNL) and SPS (CERN) at significantly higher energy values. The results obtained and the NE layers, preserved in the BECQUEREL experiment, are unique with respect to relativistic fragmentation till the present time. They include the identification of $^{8}$Be, which indicates the observation of final states up to the minimum decay energy. In general, this fact motivated the choice of nuclear clustering as an object to study by means of the NE method in the relativistic approach.

Since early 2000s the application of the NE method has been continued in the BECQUEREL experiment at the JINR Nuclotron to study the fragmentation of light nuclei (reviews \cite{Adachi,Zarubin}). Features of isotopes $^{7,9}$Be, $^{8,10,11}$B, $^{10,11}$C, $^{12,14}$N were revealed in the probabilities of dissociation channels. The $^{9}$B $\to$ $^{8}$Be$p$ decays were identified from the invariant mass calculated under the assumption of initial momentum conservation. It has been shown that the NE resolution is necessary and sufficient. The selection of $^{8}$Be is limited to 0.2 MeV (Figs. 1a and b), and $^{9}$B is limited to 0.5 MeV (Fig. 1c).

The $^{8}$Be and $^{9}$B identification became the basis to search for HS decays in the $^{12}$C $\to$ 3$\alpha$ dissociation (Fig. 1d), where the invariant mass of 3$\alpha$ triplets was limited to 0.7 MeV. The choice of these three conditions as ``cutoffs from above'' was sufficient, since the decay energy values of these three states were noticeably lower than the nearest excitations with the same nucleon composition, and the reflection of more complex excitations was small for these nuclei.

The analysis of ``white'' stars $^{12}$C $\to$ 3$\alpha$ and $^{16}$O $\to$ 4$\alpha$, not accompanied by target fragments, made it possible to establish that the fraction of events containing $^{8}$Be (HS) decays was 45 $\pm$ 4\% (11 $\pm$ 3\%) for $^{12}$C and 62 $\pm$ 3\% (22 $\pm$ 2\%) for $^{16}$O (Fig. 4d). It can be seen that the growth of 2$\alpha$- and 3$\alpha$-combinations enhances the contribution of $^{8}$Be and HS. This observation is worth to be verified for heavier nuclei, when the $\alpha$-combinatory grows rapidly with the mass number.

The simple decay selection has become possible due to the fact that the decay energy values of these three states are noticeably lower than the nearest excitations with the same nucleon composition, and the reflection of more complex excitations is small. The same approach can be applied to the further search for states immediately above the $\alpha$-particle binding thresholds. The opportunity of HS appearance via the $\alpha$-decay of $^{16}$O(0$^+_6$) has been investigated. The distribution of ``white' stars $^{16}$O $\to$ 4$\alpha$ with respect to the invariant mass of 4$\alpha$-quartets $Q_{4\alpha}$ (Fig. 2) has been mainly described by the Rayleigh distribution with the parameter $\sigma_{Q4\alpha}$ = (6.1 $\pm$ 0.2) MeV. The condition $Q_{3\alpha}$(HS) $<$ 700 keV shifts the $Q_{4\alpha}$ distribution to the low energy side. The enlarged view of the $Q_{4\alpha}$ distribution shown in the inset of Fig. 2a has indicated 9 events satisfying $Q_{4\alpha}$ $<$ 1 MeV and having the average value $\left\langle Q_{4\alpha} \right\rangle$ (RMS) = 624 $\pm$ 84(252) keV. Then the estimate of the contribution of $^{16}$O(0$^+_6$) $\to$ $\alpha$ + HS decays is 1.4 $\pm$ 0.5\% when normalized to $N_{ws}$($^{16}$O) and 7 $\pm$ 2\% when normalized to $N_{HS}$($^{16}$O).

\begin{figure}[t]
	\centerline{\includegraphics*[width=1.0\linewidth]{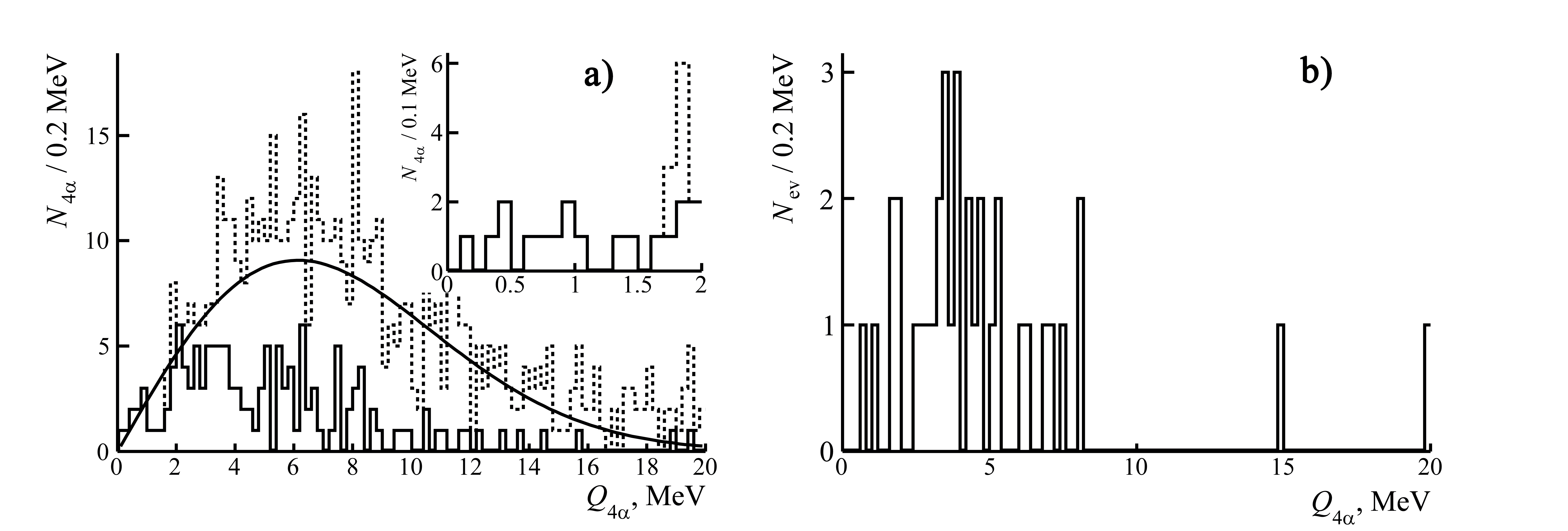}}
	\caption{Invariant mass distribution $Q_{4\alpha}$ \cite{Artemenkov} in 641 ``white'' star $^{16}$O $\to$ 4$\alpha$ at 3.65 GeV/nucleon of all 4$\alpha$-quartets (a, dots), $\alpha$HS events (a, solid) and $^{16}$O $\to$ 2$^{8}$Be (b); smooth line - Rayleigh distribution; the inset shows an enlarged part of $Q_{3\alpha}$ $<$ 2 MeV.}
\end{figure}

33 events $^{16}$O $\to$ 2$^{8}$Be have been identified, which is 5 $\pm$ 1\% of the ``white stars'' $^{16}$O $\to$ 4$\alpha$. Then the statistics $^{16}$O $\to$ 2$^{8}$Be and $^{16}$O $\to$ $\alpha$HS has the ratio of 0.22 $\pm$ 0.02. The distribution over the invariant mass $Q_{4\alpha}$ of the $^{16}$O $\to$ 2$^{8}$Be events shown in Figs. 2b points to two candidates $^{16}$O(0$^+_6$) $\to$ 2$^{8}$Be in the $Q_{4\alpha}$ $<$ 1.0 MeV region. Thus, the estimate of the probability ratio of the $^{16}$O(0$^+_6$) $\to$ 2$^{8}$Be and $^{16}$O($0^+_6$) $\to$ $\alpha$HS channels is 0.22 $\pm$ 0.17. It can be concluded that although direct dissociation dominates in the formation of HS, the search for its 4$\alpha$ ``precursor'' is possible. Since the increase in the statistics of $^{16}$O $\to$ 4$\alpha$ events has been exhausted, the study of $\alpha$-ensembles is continued for heavier nuclei.

\section*{Unstable states in dissociation of heavy nuclei}
\noindent
It seems unlikely that the above states are universal though the exotic part of the structure of the studied nuclei. An alternative is formation of $^{8}$Be nuclei in the interaction of pairs of already formed $\alpha$-particles. Then the subsequent pickup of other $\alpha$-particles and nucleons by the $^{8}$Be nuclei becomes possible. Thus, with an increase in the multiplicity of $\alpha$-particles $n\alpha$, one should expect increasing in the yield of $^{8}$Be and, possibly, $^{9}$B and HS. In such a scenario, sequential generation of $n\alpha$BEC states is probable. On the contrary, in the first variant, the inverse correlation can be expected: the increase in $n\alpha$ would result in $^{8}$Be deficit.

In this context, we have analyzed the measurements of interactions of $^{16}$O, $^{22}$Ne, $^{28}$Si, and $^{197}$Au nuclei of the Emulsion Collaboration at the JINR Synchrophasotron and the EMU Collaboration at the AGS (BNL) \cite{Zaitsev} obtained by scanning along tracks, i.e. no sampling. Wide coverage in $n\alpha$ is ensured by measurements of 1316 inelastic interactions of $^{197}$Au at 10.7 GeV/nucleon, where the fraction of events with $n\alpha$ $>$ 3 in which was 16\%. Due to the complexity of measurements, the selection condition $Q_{2\alpha}$($^{8}$Be) was weakened to $\leq$ 0.4 MeV. It turned out that the ratios of the number of events $N_{n\alpha}$($^{8}$Be) with, at least, one identified $^{8}$Be decay to their number $N_{n\alpha}$ has shown a strong increase with $n\alpha$.

The $^{197}$Au interactions contain the triplets $Q_{2\alpha p}$($^{9}$B) $\leq$ 0.5 MeV and $Q_{3\alpha}$(HS) $\leq$ 0.7 MeV. The ratio of the number of events $N_{n\alpha}$($^{9}$B), $N_{n\alpha}$(HS), and $N_{n\alpha}$(2$^{8}$Be) to $N_{n\alpha}$($^{8}$Be) has not shown a noticeable change with $n\alpha$, indicating the increase relative to $N_{n\alpha}$. However, little statistics allows us to mention only this trend. The summation of statistics $N_{n\alpha}$($^{9}$B), $N_{n\alpha}$(HS), and $N_{n\alpha}$(2$^{8}$Be) over the multiplicity $n\alpha$ and normalization to the sum $N_{n\alpha}$($^{8}$Be) leads to relative contributions of 25 $\pm$ 4\%, 6 $\pm$ 2\%, and 10 $\pm$ 2\%, respectively. The $Q_{4\alpha}$ distribution points to 4$\alpha$-quartets almost at the very threshold, in which the HS and 2$^{8}$Be decays are reconstructed under the condition of $Q_{2\alpha}$($^{8}$Be) $\leq$ 0.2 MeV. One of them with $Q_{4\alpha}$ = 1.0 (16$\alpha$, HS) MeV serves as a reference point to search for 4$\alpha$BEC.

\begin{figure}[t]
	\centerline{\includegraphics*[width=1.0\linewidth]{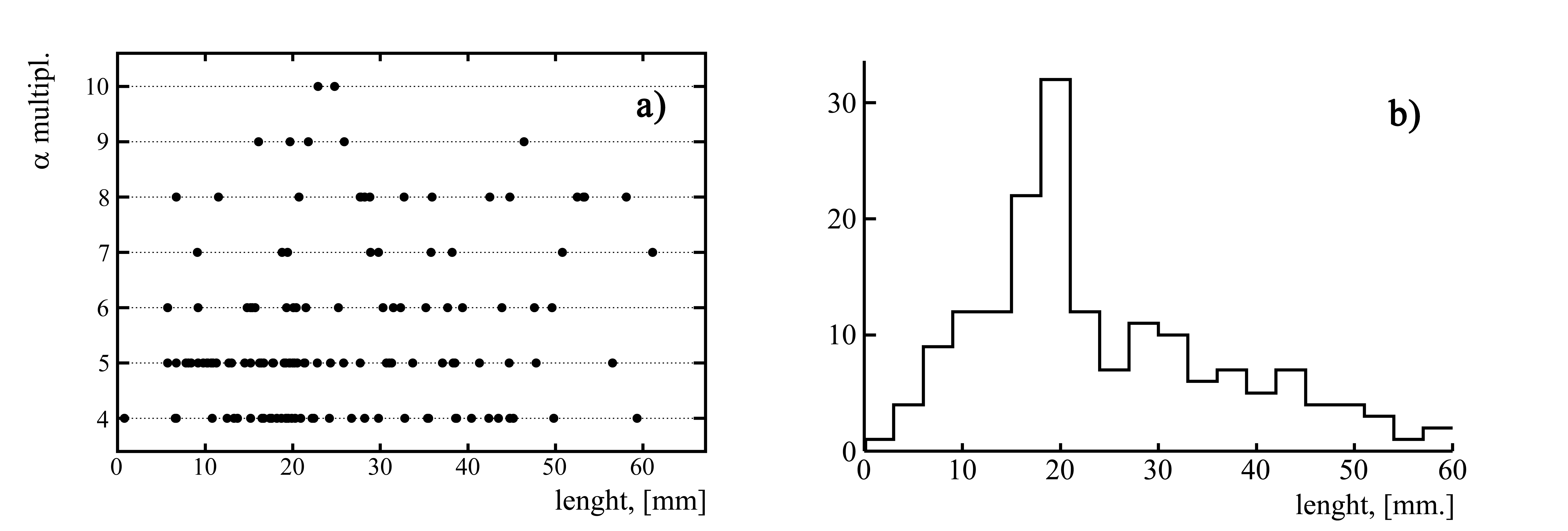}}
	\caption{Multiplicity distribution of produced $\alpha$-particles over the vertex longitudinal coordinate (a); distribution of events over the Kr nucleus range to interaction vertex (b).}
\end{figure}

\begin{figure}[t]
	\centerline{\includegraphics*[width=1.0\linewidth]{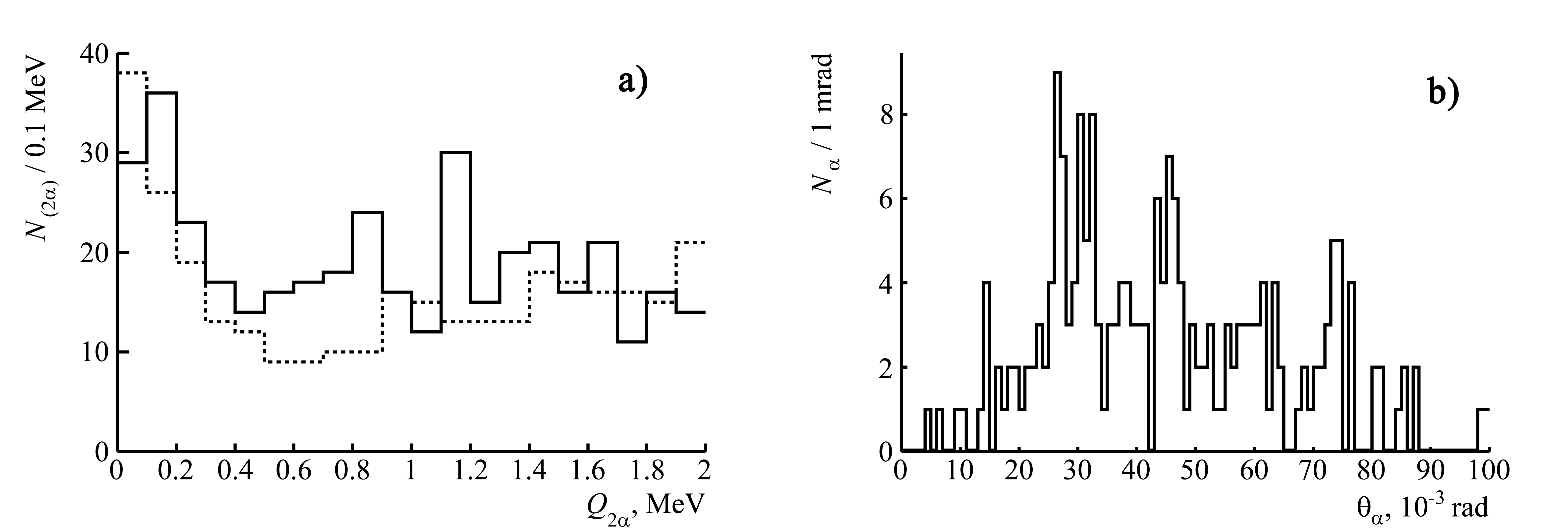}}
	\caption{Distribution of all combinations of pairs of $\alpha$-particles produced in fragmentation of $^{84}$Kr nuclei over invariant mass $Q_{2\alpha}$ $<$ 2 MeV (a) according to new measurements and early data \cite{Krasnov} (dots) and distributions of tracks of $\alpha$-particles $Q_{2\alpha}$($^{8}$Be) $\leq$ 0.4 MeV over polar angle (b).}
\end{figure}

At present, the statistics of $n\alpha$ ensembles has been increased by transverse scanning of NE layers to $^{84}$Kr nuclei at 950 MeV/nucleon (GSI, early 90s) \cite{Krasnov}. The distribution over $n\alpha$ and the longitudinal coordinate of the found event vertices is shown in Fig. 3. In this case, the energy losses up to 6 cm are approximately uniform and amount to about 9 MeV/mm (the nucleus total run is about 8 cm) \cite{Ziegler}. This effect is taken into account on the nucleus run before each interaction with the corresponding decrease in the momenta of $\alpha$-particles in the calculation of $Q_{(2-4)\alpha}$. In addition, the momentum of the fragments is taken with a factor of 0.8. Being unprincipled to select $Q_{2\alpha}$($^{8}$Be) $\leq$ 0.4 MeV, further it allows us to keep the selection condition $Q_{3\alpha}$(HS) $<$ 0.7 MeV, focusing on the $Q_{3\alpha}$(HS) peak.

Fig. 4a shows the $Q_{2\alpha}$ distribution of 173 measured stars $n\alpha$ $>$ 3. For the best selection of $^{8}$Be decays emission angles in this sample were determined from averaged values of 5-fold measurements of coordinates of 5 points of each $\alpha$-particle track at a distance to 500 $\mu$m from the vertex. The $Q_{2\alpha}$ values of 184 stars $n\alpha$ $>$ 3 from the total number $N_{ev}$ = 875 interactions of $^{84}$Kr nuclei at 950–800 MeV/nucleon are added to this distribution \cite{Krasnov}. Due to the lack of information about the position of the vertices, the energy of 875 MeV/nucleon has been assumed, and the coefficient 0.8 has not been used. These moments are not critical to identify $Q_{2\alpha}$($^{8}$Be) $\leq$ 0.4 MeV. Figure 4b shows polar angle distributions of the $\alpha$-particle tracks $Q_{2\alpha}$($^{8}$Be) $\leq$ 0.4 MeV with respect to the Kr nucleus track directions.

The statistics $N_{n\alpha}$ of the stars $n\alpha$ $>$ 3, according to new measurements and to the data of \cite{Krasnov}, is close in both cases. The similarity of the distributions of $N_{n\alpha}$ over $n\alpha$ in the margins of statistical errors indicates the correctness of the transverse scanning. The ratio $N_{n\alpha}$/$N_{ev}$ according to the data \cite{Krasnov} gives an idea of the contribution of stars $n\alpha$ $>$ 3 to the cross section for interactions with the NE composition nuclei. The statistics of $n\alpha$ $>$ 3 stars with at least one or two $^{8}$Be and HS decays is given in the Table. The statistics of the both samples has been summarized in relation to $N_{n\alpha}$($\geq$ 1$^{8}$Be)/$N_{n\alpha}$. It can be concluded that the universal effect of increasing the probability to detect $^{8}$Be in the event with $n\alpha$ increasing it manifests itself for one more nucleus and at the lowest energy value.

\begin{table}[]
	\caption{Statistics of $N_{n\alpha}$ stars $n\alpha$ $>$ 3; statistics of the sample \cite{Zaitsev} is in parentheses.}
	\begin{center}\begin{tabular}{|c| |c| |c| |c| |c| |c| |c|} \hline 
			$n\alpha$ & 4 & 5 & 6 & 7 & 8 & 9-13 \\ \hline
			$N_{n\alpha}$ & 40(69) & 50(54) & 21(27) & 10(19) & 15(12) & 7(3) \\
			$N_{n\alpha}/N_{ev}$,\% & 7.9$\pm$ 1.0 & 6.2 $\pm$ 0.9 & 3.1 $\pm$ 0.6 & 2.2 $\pm$ 0.5 & 1.4 $\pm$ 0.4 & 0.4 $\pm$ 0.2\\
			$N_{n\alpha}$($\geq$ 1$^{8}$Be) & 5(15) & 16(10) & 12(13) & 4(10) & 11(8) & 4(3) \\
			$N_{n\alpha}$($\geq$ 1$^{8}$Be)/$N_{n\alpha}$, \% & 19 $\pm$ 5 & 25 $\pm$ 6 & 52 $\pm$ 13 & 48 $\pm$ 16 & 70 $\pm$ 21 & 70 $\pm$ 35 \\
			$N_{n\alpha}$(2$^{8}$Be) & 0 & 2 & 2 & 1 & 5 & 2 \\
			$N_{HS}$ & 1 & 2 & 1 & 1 & 2 & 2 \\ \hline  
		\end{tabular}
	\end{center}
\end{table}

The new measurements made it possible to identify 12 2$^{8}$Be and 9 HS decays (Table I). The $Q_{3\alpha}$ distribution up to 2 MeV (Fig. 5a) indicates the expected concentration of $\alpha$-triplets near the HS decay energy. Taking this fact as a calibration, the correction factor of 0.8 for the momenta of $\alpha$-particles has been also included in the calculations of $Q_{4\alpha}$. Figure 5b shows polar angle distributions of $\alpha$-particle tracks $Q_{3\alpha}$(HS) $\leq$ 0.4 MeV with respect to the Kr nucleus track directions.

\begin{figure}[t]
	\centerline{\includegraphics*[width=1.0\linewidth]{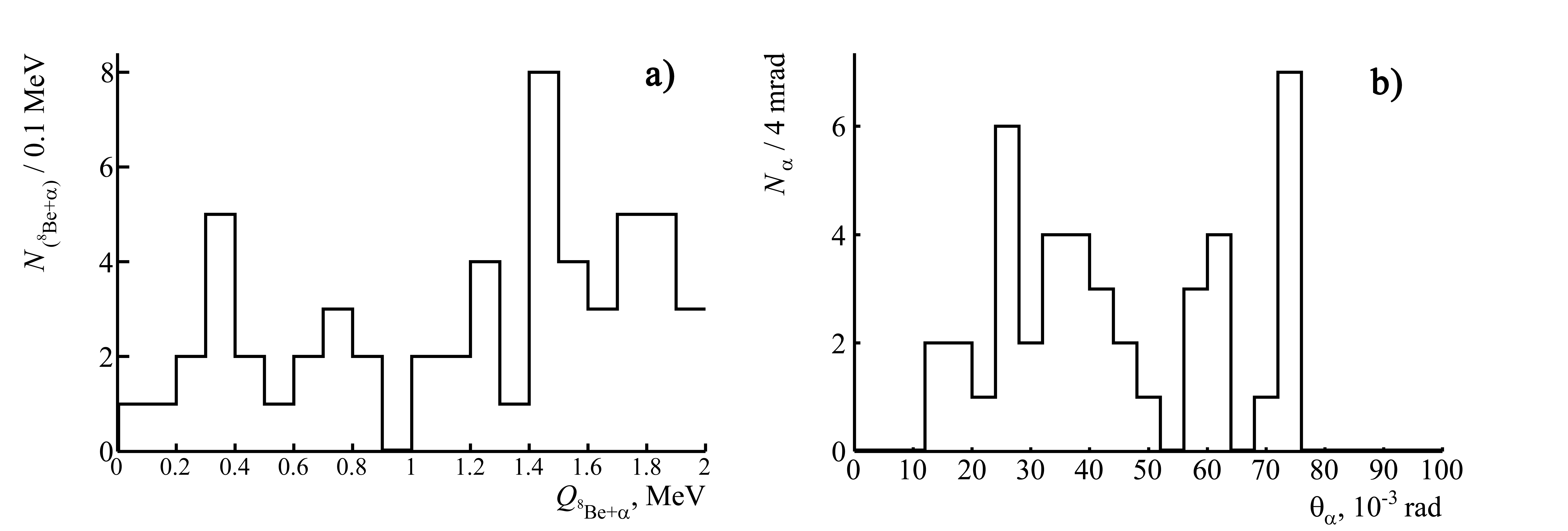}}
	\caption{Distribution of all combinations of $\alpha$-particle triplets produced in the fragmentation of $^{84}$Kr nuclei over the invariant mass $Q_{3\alpha}$ $<$ 2 MeV under the condition $Q_{2\alpha}$($^{8}$Be) $\leq$ 0.4 MeV (a) and distribution over the polar angle of $\alpha$-particle tracks $Q_{3\alpha}$(HS) $\leq$ 0.7 MeV and $Q_{2\alpha}$($^{8}$Be) $\leq$ 0.4 MeV (b).}
\end{figure}

The $Q_{4\alpha}$ distributions up to 10 MeV are shown in Figs. 6 under the conditions $Q_{2\alpha}$($^{8}$Be) $\leq$ 0.4 MeV and $Q_{3\alpha]}$(HS) $<$ 0.7 MeV (a), as well as the condition for two pairs of $\alpha$-particles $Q_{2\alpha}$($^{8}$Be) $\leq$ 0.4 MeV (b). The both distributions indicate a 4$\alpha$ quartet at $n\alpha$ = 6 with the isolated value $Q_{4\alpha}$ = 0.6 MeV, corresponding to both $\alpha$HS and 2$^{8}$Be variants. The energy of the Kr nucleus, considering the correction in this event, is equal to 700 MeV/nucleon, and the polar angles with the respect to the direction of the Kr trace in the $\alpha$-particle quartet are: 58, 63, 73, and 75 10$^{-3}$ rad, respectively. Not contradicting the $^{16}$O(0$^+_6$) decay, this single observation serves as a starting point to further accumulate the statistics on the 4$\alpha$BEC problem.

\begin{figure}[t]
	\centerline{\includegraphics*[width=1.0\linewidth]{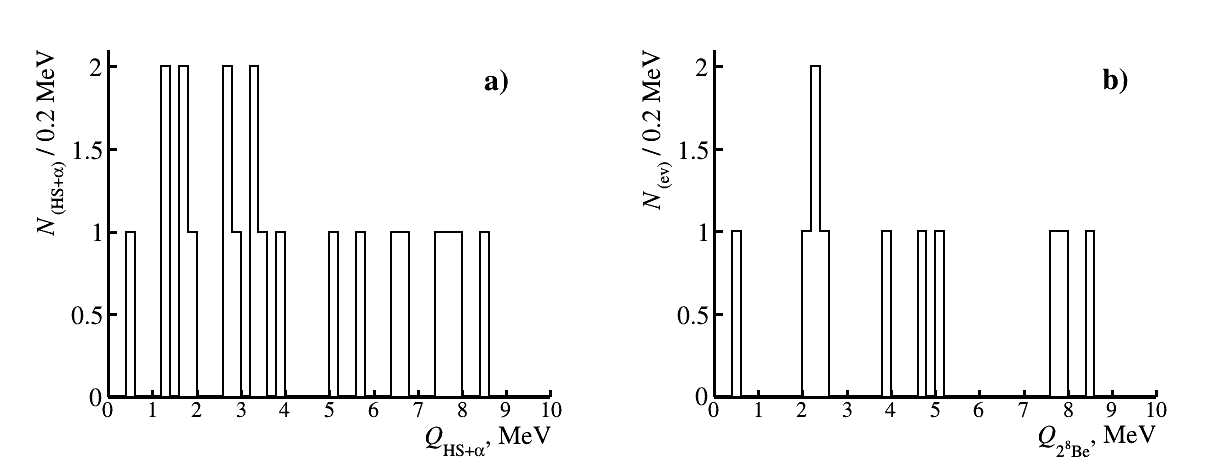}}
	\caption{Distribution of all combinations of quartets of $\alpha$-particles produced in the fragmentation of $^{84}$Kr nuclei over invariant mass $Q_{3\alpha}$ $<$ 10 MeV under conditions $Q_{2\alpha}$($^{8}$Be) $\leq$ 0.4 MeV and $Q_{3\alpha}$(HS) $<$ 0.7 MeV (a) and condition $Q_{2\alpha}$($^{8}$Be) $\leq$ 0.4 MeV (b) for two pairs of $\alpha$-particles.}
\end{figure}

\section*{Isobar analog states in light nuclei}
\noindent
The study of unstable states has shown the opportunity of searching for more complex excitations in light nuclei - isobar-analogue states (IAS), indicating the rearrangement in the direction of similarity with less stable isobars having less $\alpha$-clustering. Despite the high energy (13-18 MeV), IASs are distinguished by widths $\Gamma$ smaller than those of neighboring excitations, associated with the prohibition of their decays along the isospin $\Delta$T = 1, i.e., with increasing in $\alpha$-clustering. It can be assumed that the IAS in light nuclei excite the $hn$ and $tp$ configurations with T = 1 up to the coupling threshold (Fig. 7). Here the $^{3}$He cluster is denoted as $h$ (helion). Such a virtual transition may be the result of a nucleon spin flip in $\alpha$-quartets $nn-pp$ (Fig. 7a). Being impossible in a free $\alpha$-particle, it can affect the diffraction scattering of $\alpha$-particles.

The manifestation of $hn-tp$ pairs can be traced starting from $^{8}$Be (Fig. 7b), in which there is a doublet of excitations $^{8}$Be(16.6) with a width of $\Gamma$ = 108 keV and $^{8}$Be(16.9) with $\Gamma$ = 74 keV, mixed in isospin T = 0 + 1. Being located below the threshold $^{7}$Li + $p$ (17.255) and decomposing only into an $\alpha$-pair, these levels are candidates for the $\alpha$ + ($hn/tp$) configuration. The $^{8}$Be(16.6 + 16.9) levels are quite far from the nearest $^{8}$Be$_{4^{+}}$(11.4) excitation with $\Gamma$ = 3.5 MeV, which allows their joint identification in the relativistic fragmentation $^{9}$Be $\to$ 2$\alpha$. Above them, there is the IAS $^{8}$Be (17.640) with T = 1 and $\Gamma$ = 10.7 keV above the threshold of the isospin-allowed $^{7}$Li + $p$ decay. Due to the difference in the magnetic rigidity of the decay products, identification of the latter is convenient in the electronic experiment.

\begin{figure}[t]
	\centerline{\includegraphics*[width=1.0\linewidth]{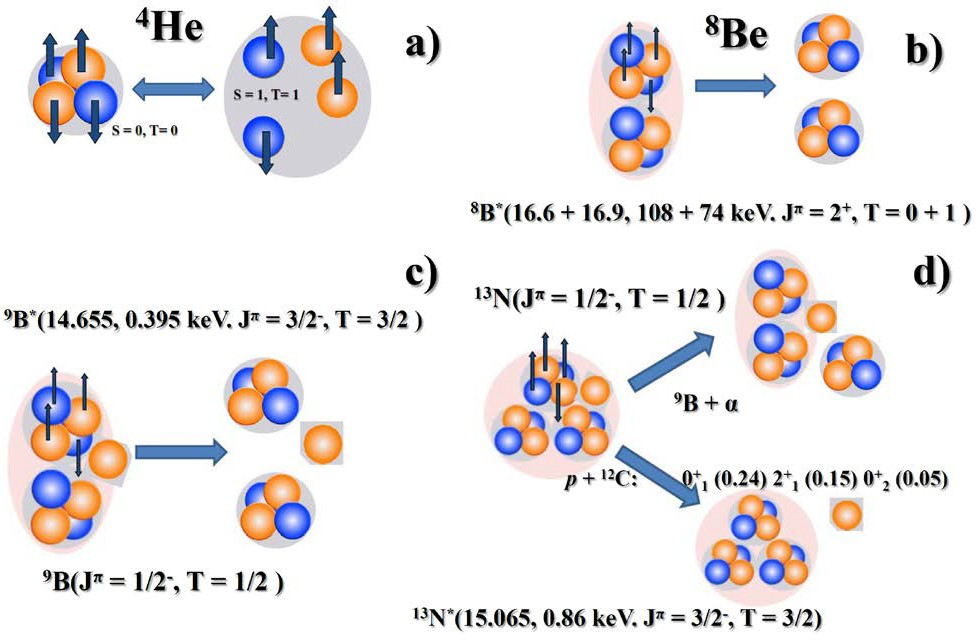}}
	\caption{Scenario for occurrence of IAS in light nuclei based on perturbation of $\alpha$-particle configuration (a) in $^{8}$Be (b), $^{9}$B (c) and $^{13}$N (d).}
\end{figure}

The addition of a proton leads to the excitation $\alpha$ + ($hn/tp$) + $p$ with T = 3/2 (Fig. 7c), which could correspond to IAS $^{9}$B(14.655) with width $\Gamma$ = 0.395 keV. When studying the coherent dissociation of $^{10}$C nuclei at 2 GeV/nucleon, the leading role of the 2He2H channel (82\%) has been found to be largely due to $^{9}$B decays (30\%) (review \cite{Adachi}). Complete coincidence of decays of the ground states $^{9}$B $\to$ $^{8}$Be has appeared which makes $^{10}$C an efficient source of $^{9}$B. Available angular measurements in ``white'' stars $^{10}$C $\to$ 2$\alpha$2$p$ enable one to check the presence of $^{9}$B(14.655) decays in them. They are supplemented by measurements of 2$\alpha$2$p$ stars containing target fragments or produced mesons, whose statistics is increasing.

Currently, the accelerated search for $^{14}$N $\to$ 3$\alpha$(+H) events at 2 GeV/nucleon is carried out on the transverse scanning of NE layers. Previously, the leadership of the 3HeH channel in the distribution over the $^{14}$N charge-conserving fragmentation channels has been established, and the contribution of $^{8}$Be $\to$ 2$\alpha$ decays of 25$–$30\% has been revealed \cite{Zarubin,Schedrina}. The initial goal is to determine the contributions $^{8}$Be, $^{9}$B, and HS. Since $^{14}$N fragmentation is the source of 3$\alpha p$ ensembles, another object is the $^{13}$N(15.065) IAS with isospin T = 3/2 in the $^{13}$N excitation spectrum at 5.6 MeV above the $^{9}$B$\alpha$ threshold. As a consequence of the isospin prohibition, the width of $^{13}$N$^*$(15.065) is only $\Gamma$ = 0.86 keV. Basically, the decays $^{12}$C(0$^+_2$)$p$ and $9$B$\alpha$, which have probabilities of 5\% each, can serve as signatures of $^{13}$N(15.065) \cite{Adelberger}.

Consider $^{13}$N(15.065) in the $\alpha$-cluster pattern (Fig. 7d). The values T = 3/2 and J = 3/2 are possible in the configurations 2$\alpha$+($hn$)+$p$ and 2$\alpha$+($tp$)+$p$ involving virtual pairs $hn$ or $tp$ with spin J = 1. The transition is possible upon spin flip $S$-wave nucleon in the 3$\alpha p$ ensemble, without completely overcoming the coupling threshold $hn$ and $tp$ (about 20 MeV). The $^{13}$N(15.065) decay is initiated by the return of a nucleon to the $\alpha$-cluster, and the released energy is realized via emission of a proton or $\alpha$-particle and the excited and ground states $^{12}$C and $^{9}$B, respectively. As a signal of the IAS branch, the detection of $^{14}$N $\to$ $^{13}$N(15.065) would motivate their search for neighboring nucleus fragmentation. Another opportunity is to search for the $^{14}$N($>$ 20.4 MeV) T = 1 state from the $\alpha d$ decays, also suppressed by isospin.

Fig. 8 presents the analysis status. It indicates the probable presence of IAS in the range from 5-9 MeV above the 3$\alpha p$ threshold, which is satisfactory in this approach. Thus, a bound $\alpha$-particle manifests itself as an elastically deformable object which underlies the whole family of fairly long-lived states. Its relaxation to the $S$-state determines the final states of IAS decays.

\begin{figure}[t]
	\centerline{\includegraphics*[width=0.6\linewidth]{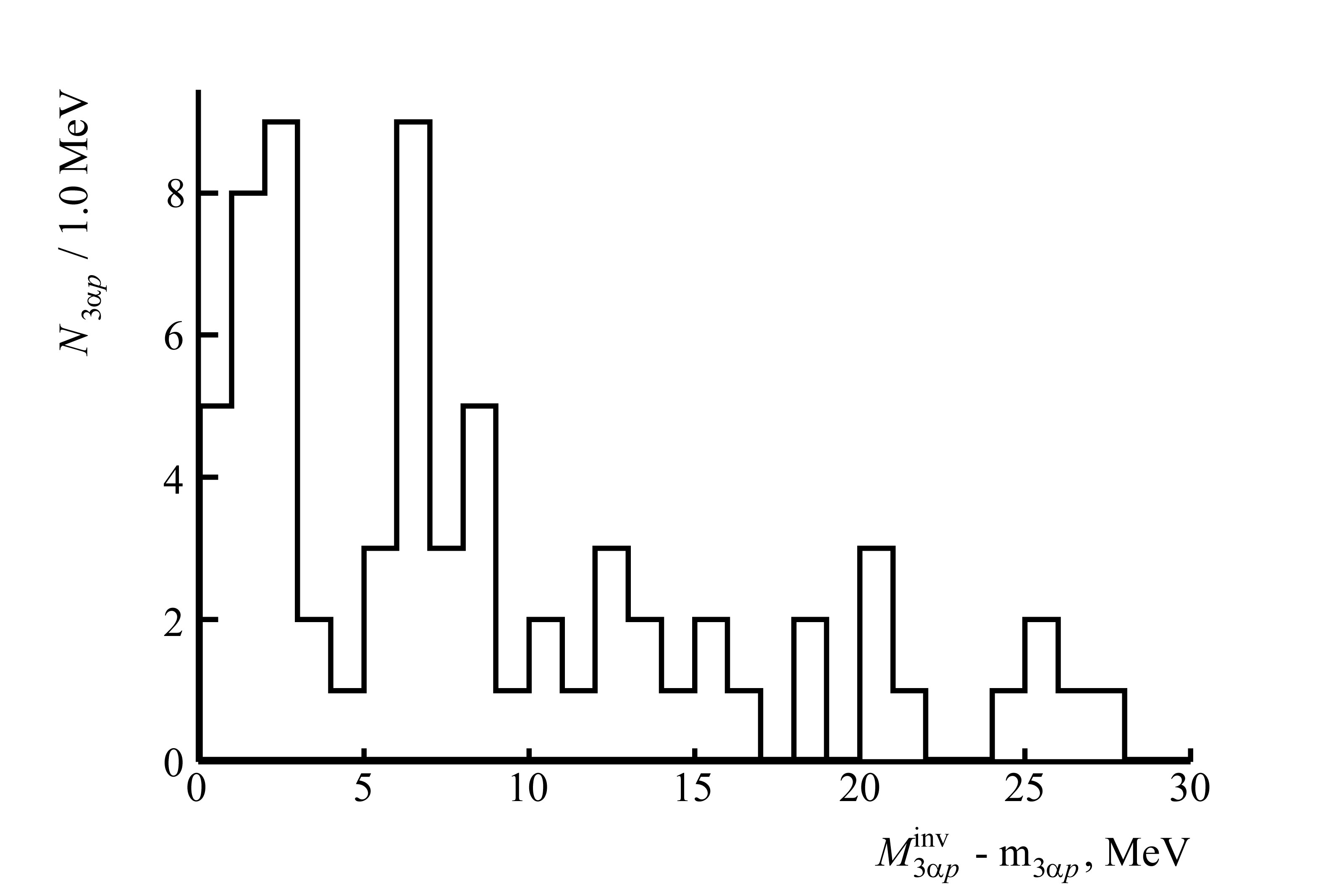}}
	\caption{Distributions of 80 events $^{14}$N $\to$ 3$\alpha p$ over invariant mass $Q_{3\alpha p}$.}
\end{figure}

\section*{Conclusions}
\noindent
The status of the BECQUEREL experiment aimed at solving topical problems in the physics of nuclear clusters has been presented. Due to the unique sensitivity and spatial resolution, the nuclear emulsion method makes it possible in a unified approach to study multiple final states arising in the dissociation of the widest variety of nuclei.

Currently, the research is focused on the concept of $\alpha$-particle Bose-Einstein condensate ($\alpha$BEC), the extremely cold state of several $S$-wave $\alpha$-particles near the coupling thresholds. The unstable $^{8}$Be nucleus has been described as 2$\alpha$BEC, and the $^{12}$C(0$^+_2$) excitation or Hoyle state (HS) as 3$\alpha$BEC. Decays $^{8}$Be $\to$ 2$\alpha$ and $^{12}$C(0$^+_2$) $\to$ $^{8}$Be$\alpha$ can serve as signatures for more complex $n\alpha$BEC decays. Thus, the 0$^+_6$ state of the $^{16}$O nucleus at 660 keV above the 4$\alpha$ threshold, considered as 4$\alpha$BEC, can sequentially decay $^{16}$O(0$^+_6$) $\to$ $\alpha$$^{12}$C(0$^+_2$) or $^{16}$O(0$^+_2$) $\to$ 2$^{8}$Be. Its search is carried out in several experiments on the fragmentation of light nuclei at low energies. Confirmation of the existence of this and more complex forms of $\alpha$BEC could provide a basis to expand scenarios for the fusion of medium and heavy nuclei in nuclear astrophysics.

The consideration of $\alpha$BEC as an invariant phenomenon indicates the opportunity of searching for it in nuclear emulsion layers longitudinally exposed to relativistic nuclei. In this case, the invariant mass of ensembles of He and H fragments can be determined from the emission angles in the approximation of conservation of momentum per nucleon of the parent nucleus. Owing to the extremely small energies and widths, the decays $^{8}$Be and HS, as well as $^{9}$B $\to$ $^{8}$Be$p$, have been identified in the fragmentation of light nuclei by the upper constraint on the invariant mass.

After being tested, this approach has been used to identify $^{8}$Be and HS and search for more complex $n\alpha$BEC states in the fragmentation of medium and heavy nuclei. Recently, based on the statistics of dozens of $^{8}$Be decays, the increased probability of detecting $^{8}$Be in the event with the increasing in the number of relativistic $\alpha$-particles has been found. A preliminary conclusion is drawn that the contributions from $^{9}$B and HS decays also increase. The exotically large sizes and lifetimes of $^{8}$Be and HS assume an opportunity of synthesizing $\alpha$BEC by successively connecting the emerging $\alpha$ particles 2$\alpha$ $\to$ $^{8}$Be, $^{8}$Be$\alpha$ $\to$ $^{12}$C(0$^+_2$), $^{12}$C(0$^+_2$)$\alpha$ $\to$ $^{16}$O(0$^+_6$), 2$^{8}$Be $\to$ $^{16}$O(0$^+_6$) and further with a probability decreasing at each step, when $\gamma$-quanta or recoil particles are emitted.

The main task of the forthcoming stage is to elucidate the connection between the appearance of $^{8}$Be and HS and the multiplicity of $\alpha$-ensembles and search on this basis for the decays of the $^{16}$O(0$^+_6$) state. In this regard, the BECQUEREL experiment aims at measuring multiple channels of $^{84}$Kr fragmentation up to 1 GeV/nucleon. There is a sufficient number of NTE layers, the transverse scanning of which makes it possible to achieve the required statistics. The presented data are the first contribution to the targeted search for 4$\alpha$BEC. Although the data obtained are encouraging, the fold increase in statistics is required to confirm 4$\alpha$BEC.

The search for decays of isobar-analogue states (IAS) has begun to continue the study of the light nucleus fragmentation. Manifesting at high excitation energy, but also having very small widths, IASs serve as ``firehouses'' to structurally rearrange in the direction of similarity with their fewer stable isobars. In this context, the analysis of nuclear emission irradiation with $^{9}$Be, $^{14}$N, $^{22}$Ne, $^{24}$Mg, $^{28}$Si nuclei has been resumed.

The solution of the tasks is feasible with the motorized microscope Olympus BX63, recently supplied for the BECQUEREL experiment. Mastering its capabilities is a special methodological challenge. Hopefully, its application and advantages in the image analysis will make it possible to give the entirely new dimension of using the NE method.

\end{document}